\documentclass[aps,prb,twocolumn,floatfix,showpacs,superscriptaddress]{revtex4}
\usepackage[dvips]{color}
\usepackage{graphicx}
\usepackage{amsmath,amssymb,bm}
\usepackage[hypertex,breaklinks=true,colorlinks=false]{hyperref}
\usepackage{color}

\newcommand{\bra}[1]{\langle #1|}
\newcommand{\ket}[1]{|#1 \rangle}

\newcommand{\ua}{\uparrow}
\newcommand{\da}{\downarrow}

\def\rnum#1{\expandafter{%
\romannumeral #1}}
\def\Rnum#1{\uppercase\expandafter{%
\romannumeral #1}}

\begin{document}

\title{
Unconventional N\'eel and dimer orders in a spin-$\frac12$ frustrated ferromagnetic chain with easy-plane anisotropy 
}
\author{Shunsuke Furukawa}
\affiliation{Department of Physics, University of Toronto, Toronto, Ontario, Canada M5S 1A7}
\affiliation{Condensed Matter Theory Laboratory, RIKEN, Wako, Saitama 351-0198, Japan}
\author{Masahiro Sato}
\affiliation{Condensed Matter Theory Laboratory, RIKEN, Wako, Saitama 351-0198, Japan}
\author{Akira Furusaki}
\affiliation{Condensed Matter Theory Laboratory, RIKEN, Wako, Saitama 351-0198, Japan}
\date{\today}


\begin{abstract}
We study the ground-state phase diagram of 
a one-dimensional spin-$\frac12$ easy-plane XXZ model
with a ferromagnetic nearest-neighbor (NN) coupling $J_1$
and a competing next-nearest-neighbor (NNN) antiferromagnetic coupling $J_2$
in the parameter range $0<J_2/|J_1|<0.4$.
When $J_2/|J_1|\lesssim1/4$,
the model is in a Tomonaga-Luttinger liquid phase
which is adiabatically connected to the critical phase
of the XXZ model of $J_2=0$.
On the basis of the effective (sine-Gordon) theory and numerical analyses
of low-lying energy levels of finite-size systems,
we show that the NNN coupling induces phase transitions
from the Tomonaga-Luttinger liquid to gapped phases 
with either N\'eel or dimer order.
Interestingly, these two types of ordered phases appear alternately
as the easy-plane anisotropy is changed towards the isotropic limit. 
The appearance of the antiferromagnetic (N\'eel) order in this model is
remarkable, as it is strongly unfavored by both the easy-plane ferromagnetic
NN coupling and antiferromagnetic NNN coupling in the classical-spin picture. 
We argue that emergent trimer degrees of freedom 
play a crucial role in the formation of the N\'eel order.
%
\end{abstract}
\maketitle


\section{Introduction}

The search for novel orders 
arising from geometric frustration and quantum fluctuations in low-dimensional
quantum spin systems 
has been a subject of intensive theoretical and experimental research.
One-dimensional (1D) systems offer particularly ideal grounds for
theoretical studies, as powerful non-perturbative methods are available
(for a review, see, for example, Ref.~\onlinecite{Lecheminant05}).  
A prototypical example hosting a variety of fascinating phenomena 
is the spin-$\frac12$ frustrated chain 
with the nearest-neighbor (NN) and next-nearest-neighbor (NNN) exchange
interactions: 
\begin{equation} \label{eq:H}
  H= \sum_{m=1}^2\sum_{j=1}^{L} J_m
  \left( 
     S^x_j S^x_{j+m} + S^y_j S^y_{j+m} + \Delta S^z_j S^z_{j+m}
  \right) .
\end{equation}
Here, $\bm{S}_j=(S^x_j,S^y_j,S^z_j)$ is a spin-$\frac12$ operator at the site $j$,
and $\Delta$ is an XXZ anisotropy. 
The model has frustration as long as the NNN coupling $J_2$ is
antiferromagnetic, 
irrespective of the sign of $J_1$. 

Much of recent interest in the model (\ref{eq:H}) has
been focused
on the case when ferromagnetic (FM) $J_1<0$ and antiferromagnetic (AFM) $J_2>0$ compete. 
The growing interest was triggered by experimental studies of various
quasi-1D edge-sharing cuprates\cite{Hase04,Masuda05,Enderle05,Drechsler07} 
which can be modelled by the Hamiltonian (\ref{eq:H}).
Theoretical studies
have uncovered a rich phase diagram of the model in a magnetic field,
which includes the vector chiral phase, spin nematic, and various other
multipolar liquids.\cite{Chubukov91,Kolezhuk05,Heidrich06,Vekua07,Kecke07,Hikihara08,Sato09,Sudan09,Heidrich09} 
Even without a magnetic field, 
the model \eqref{eq:H} shows a variety of phases depending on
the values of the exchange anisotropy $\Delta$ 
and the frustration parameter $J_2/|J_1|$. 
Earlier studies have discussed 
the dimer order\cite{Tonegawa90,Somma01}  
and the stability of the vector chiral order\cite{FSSO08,Sirker09} in the easy-plane case $0\le\Delta<1$, 
and the partial ferromagnetism\cite{Tonegawa90} in the easy-axis case $\Delta>1$.
In this paper we show that
the phase diagram of the $J_1$-$J_2$ chain with easy-plane XXZ anisotropy, 
Eq.\ (\ref{eq:H}), 
exhibits unexpectedly complicated and interesting phase structure, 
where N\'eel and dimer ordered phases alternate more than twice, 
near a FM critical point $(J_2/|J_1|,\Delta)=(0.25,1)$. 
Our analysis is based on the combination of bosonization and numerical analysis of
finite-size energy spectrum, which has been proven successful
in the study of many 1D quantum spin models including the model (\ref{eq:H})
with the antiferromagnetic NN coupling.
The ground-state phase diagram of the anisotropic $J_1$-$J_2$ spin chain \eqref{eq:H}
is well understood 
in the {\em antiferromagnetic} case, $J_1>0$ and $J_2> 0$
(see, e.g., Refs.~\onlinecite{Majumdar69,Haldane80,Haldane82,Okamoto92,Nomura94,White96,Nersesyan98,Hikihara01}).
For small $J_2/J_1$ and with easy-plane anisotropy $\Delta < 1$, 
the model is in a gapless phase which is described
as a Tomonaga-Luttinger liquid (TLL). 
The gapless TLL phase has instabilities towards antiferromagnetic (N\'eel)
order and dimerization (as exemplified by the exact singlet dimer ground
state\cite{Majumdar69} for $J_2/J_1=1/2$ and $\Delta>-1/2$).
Both N\'eel and dimer ordered phases are gapped.
The N\'eel order is induced by easy-axis anisotropy $\Delta>1$,
while the dimer order appears when $J_2/J_1$ is larger than
some critical value ($\approx0.24$ at $\Delta=1$).
Haldane\cite{Haldane80,Haldane82} showed that the quantum phase transitions
to these ordered phases
can be understood in a unified way within the sine-Gordon (SG) model
which is obtained by bosonizing the Hamiltonian (\ref{eq:H}).
The two types of orders arise as a bosonic field is locked
at two distinct values
by the cosine potential coming from the Umklapp scattering
of Jordan-Wigner fermions.
It is also worth noting that, for large $J_2/J_1$,
a vector chiral ordered phase with gapless excitations
appears for a certain range of easy-plane
anisotropy,\cite{Nersesyan98,Hikihara01} 
while the dimer ordered phase persists in the isotropic
case $\Delta=1$.\cite{White96}




In this paper,
we apply the SG formalism of Haldane\cite{Haldane80,Haldane82}
to the frustrated {\em ferromagnetic} spin chain \eqref{eq:H}. 
Our starting point is the FM $J_1$-only chain 
with {\it easy-plane} anisotropy $0\le \Delta <1$.
In this case the system is a TLL
whose properties are understood in great detail 
from the Bethe ansatz and the bosonization. 
In particular, the parameters in the effective SG theory are
known exactly.\cite{Lukyanov98,Lukyanov03}
According to these exact results,
rather interestingly,
there occurs a series
of sign changes in
the coupling of the cosine potential, 
as a function of the anisotropy $\Delta$. 
In the $J_1$ XXZ chain, the cosine potential is irrelevant in the
renormalization-group (RG) sense
and does not make any significant impact on the long-distance properties. 
However, it is made relevant by large enough $J_2$ coupling,
and as a result we observe alternating appearance of N\'eel and dimer
ordered phases,
as the exchange anisotropy is changed towards the
FM critical point $(J_2/|J_1|,\Delta)=(0.25,1)$. 
We determine the phase boundaries accurately 
using finite-size scaling analysis of discrete energy levels,
the so-called
level spectroscopy method,\cite{Okamoto92,Nomura94,Nomura95}
which combines the SG theory with the numerical exact diagonalization. 
The appearance of the antiferromagnetic spin ordering in the $z$ direction
is particularly counterintuitive;
on the classical level, 
such spin configuration is strongly unfavored by the easy-plane FM $J_1$
coupling as well as by the AF $J_2$ coupling. 
We analyze the correlations and the reduced density matrices of
the ground state in the N\'eel phase
using the infinite time-evolving block decimation (iTEBD)
algorithm,\cite{Vidal07} a numerical method
which can directly address physical quantities in the thermodynamic limit.
We will then argue that emergent trimer degrees of freedom 
play a crucial role in the formation of the N\'eel ordered state.

The present study focuses on the region with $J_2/|J_1|\lesssim 0.4$. 
The analysis of the region with larger $J_2/|J_1|$ is also done by iTEBD 
and will be reported elsewhere.\cite{FSO}
It is found that for $J_2/|J_1|>1/4$, 
the vector chiral ordered phase with gapless excitations 
is robust up to a very weak anisotropy $\Delta\approx 1$ 
and appears in a wide range of the parameter space. 
This is in contrast to the antiferromagnetic $J_1$-$J_2$ model, 
where the vector chiral ordered phase has been identified\cite{Hikihara01} 
only for large $J_2/J_1 (\gtrsim1.2)$. 




\section{Formalism}

\subsection{Effective field theory}

We first consider the case $J_2=0$, i.e.,
the easy-plane XXZ chain with $J_1<0$.
We parametrize the exchange anisotropy as
\begin{equation}
 \Delta = \cos(\pi\eta), \quad 0<\eta \le \frac12. 
\label{anisotropy}
\end{equation}
In this parameter range, the model is in the gapless TLL phase.
In the bosonization formalism (for a review, see, e.g., Refs.~\onlinecite{Giamarchi04,Gogolin98}),  
its low-energy effective theory is the quantum SG theory defined by the Hamiltonian density
\begin{equation}\label{eq:sineGordon}
 {\cal H} =
 \frac{v}2 \left[
 K \left(\frac{d\theta}{dx}\right)^2
  + \frac1K \left(\frac{d\phi}{dx}\right)^2
 \right]
 - \frac{v\lambda}{2\pi} \cos (\sqrt{16\pi}\phi),
\end{equation}
where the bosonic field $\phi$ and its dual counterpart $\theta$ 
obey the commutation relation 
$[\phi(x),\theta(y)]=-(i/2)[1+\mathrm{sgn}(x-y)]$, 
and the lattice constant is set to unity. 
The TLL parameter $K$ and the spin velocity $v$ are given by 
\begin{equation}\label{eq:J1chain_Kv}
 K=\frac1{2\eta}, \quad 
 v=|J_1| \frac{\sin (\pi \eta)}{2(1-\eta)}.
\end{equation}
The cosine potential in Eq.~\eqref{eq:sineGordon} has
scaling dimension $4K$ and
is irrelevant in the XXZ chain
with the exchange anisotropy (\ref{anisotropy}).
We keep the cosine term because it will drive the phase
transitions from the TLL to the N\'eel and dimer ordered phases
in the presence of the frustrated NNN coupling $J_2>0$.
When $J_2=0$, the exact value of the coupling constant $\lambda$ 
is known to be\cite{Lukyanov98,Lukyanov03,Furusaki05}
\begin{equation}\label{eq:coupling_Lukyanov}
\lambda_0
  = -\frac4\pi \sin\!\left( \frac\pi\eta \right)
 \left[ \Gamma\!\left(\displaystyle \frac1\eta\right) \right]^2
 \left[\frac{ \Gamma\!\left(\displaystyle 1+\frac\eta{2-2\eta}\right) }
            {\sqrt{4\pi}\,\Gamma\!\left(\displaystyle 1+\frac1{2-2\eta}\right)}
 \right]^{\frac2\eta -2}
\!\!,
\end{equation}
where the normalization condition on the zero-temperature correlator
of vertex operators,
\begin{equation}\label{eq:CFT_norm}
 \langle e^{i\mu\phi(x)} e^{-i\mu\phi(x^\prime)} \rangle
  = |x-x^\prime|^{-K\mu^2 / 2\pi}, \quad
 |x-x^\prime|\gg1,
\end{equation}
is assumed.
Notice that $\lambda_0$ vanishes at $\eta=1/n$, i.e., 
\begin{equation}\label{eq:XXZ_special}
 \Delta= \cos\left( \pi/n \right), \quad
 n=3,4,\dots~.
\end{equation}
At these points, the XXZ model is invariant
under the action of the loop algebra $sl_2$, 
leading to some non-trivial degeneracies
in the energy spectrum.\cite{Deguchi01,Bortz09} 
We will observe one example of such degeneracies in our numerical result later.

The spin operators of the original lattice model (\ref{eq:H}) are
expressed as\cite{note_Sp}
\begin{eqnarray}
 S_j^z &=&
 \frac1{\sqrt{\pi}} \frac{d\phi}{dx}
 + (-1)^j a \cos (\sqrt{4\pi}\phi)+\dots, \label{eq:Sz_boson} \\
 S_j^+ &=&
 e^{i\sqrt{\pi}\theta} \left[ b_0
 + (-1)^j b_1 \cos(\sqrt{4\pi}\phi) +\dots \right],
\label{eq:Sp_boson}
\end{eqnarray}
where $a$, $b_0$, and $b_1$ are non-universal
constants.\cite{Lukyanov97,Lukyanov99,Hikihara98} 
Similarly, the staggered part of the NN exchange energy is
given by
\begin{equation}
\bm{S}_j\cdot\bm{S}_{j-1}-\bm{S}_j\cdot\bm{S}_{j+1}
=c(-1)^j \sin(\sqrt{4\pi}\phi)+\ldots,
\label{eq:dimer_boson}
\end{equation}
where $c$ is another non-universal constant.

Including the NNN coupling $J_2>0$ changes the parameters
($v$, $\lambda$, and $K$) in the effective
theory \eqref{eq:sineGordon}.
A perturbative calculation using \eqref{eq:Sz_boson} and \eqref{eq:Sp_boson}
yields the renormalized TLL parameter
for $J_2/|J_1|\ll 1$,
\begin{equation}
 K = K_0
 \left[
 1
 - \frac{J_2}{v_0}\left(\frac{2\pi b_1^2}{K_0} + \frac{K_0\Delta}{\pi}\right)
 \right],
\end{equation}
where $K_0$ and $v_0$ are the values of $K$ and $v$
at $J_2=0$ given in Eq.~\eqref{eq:J1chain_Kv}. 
This indicates that $K$ decreases as $J_2/|J_1|$ increases.
As long as $K$ is larger than 1/2, the cosine term is irrelevant
and the system is in the TLL phase.
As $J_2/|J_1|$ is further increased,
the cosine term in Eq.~\eqref{eq:sineGordon} will eventually become
relevant ($4K<2$).
The positive and negative values of $\lambda$ will then trigger
the N\'eel and dimer orders, respectively, as one can see
from Eqs.~(\ref{eq:Sz_boson}) and (\ref{eq:dimer_boson}).
We analyze such ordering transitions through a combination of 
the RG analysis of the SG model \eqref{eq:sineGordon} 
and the finite-size spectra obtained from numerical exact diagonalization.


\begin{figure}
\begin{center}
\includegraphics[width=0.35\textwidth]{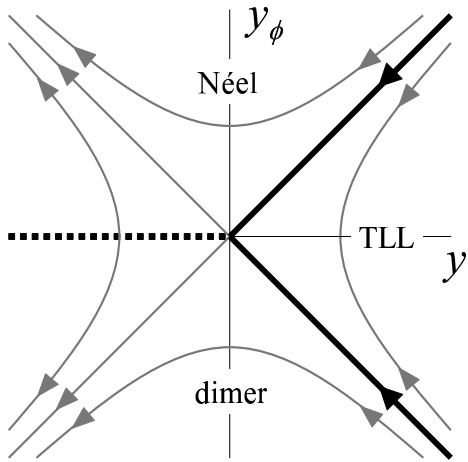}
\end{center}
\caption{
RG flow diagram of the sine-Gordon theory \eqref{eq:sineGordon}. 
The thick solid lines indicate the Kosterliz-Thouless-type
transitions\cite{Kosterlitz73} 
between the TLL and the two ordered phases. 
The dotted line indicates the Gaussian transition
between the two ordered phases. 
}
\label{fig:sineGordon}
\end{figure}

\subsection{Renormalization group flow}
We discuss the RG flows of the SG theory \eqref{eq:sineGordon}
by varying a short-distance cutoff $\alpha$.
To this end, it is convenient to parametrize 
\begin{equation}
 \alpha=\alpha_0 e^{l},\quad
 K= \frac12+\frac{y(l)}4,\quad
 \lambda = \frac{y_\phi(l)}{\alpha^2},   
\end{equation}
where $\alpha_0$ is the initial cutoff. 
Under an infinitesimal change $l\to l+dl$,
the dimensionless parameters 
$y(l)$ and $y_\phi(l)$ change according to the RG equations  
\begin{equation}
 \frac{dy (l)}{dl} = -y_\phi^2(l),\quad
 \frac{dy_\phi(l)}{dl} = -y_\phi(l) y (l).
\end{equation}
The RG flow diagram is drawn in Fig.~\ref{fig:sineGordon}. 
When $0<y<|y_\phi|$, the cosine potential finally vanishes, 
and the system is described as the TLL. 
The transitions from the TLL to the N\'eel and dimer phases 
occur along the half lines $y_\phi=+ y$ and $y_\phi=-y$
with $y>0$ (thick solid lines), respectively. 
The transition between the two ordered phases occurs
along the Gaussian fixed line $y_\phi=0$ with $y<0$ (dotted line).

\subsection{Level spectroscopy}

On the basis of the RG flow diagram of the SG model
in Fig.~\ref{fig:sineGordon}, 
Okamoto and Nomura\cite{Okamoto92,Nomura94,Nomura95} developed
a simple numerical method which allows
to determine the phase transition lines precisely. 
The key idea of this method is to look at the lowest excited states 
in the three different phases.
Under the periodic
boundary condition, 
the eigenstates of a finite-size system of length $L$
are labeled by 
the magnetization $S^z=\sum_j S_j^z$, 
the wave number $q= 2\pi k/L~(k\in \mathbb{Z})$, 
the (bond-centered) parity $P=\pm 1$, 
and the spin reversal $T=\pm 1$. 
In the parameter region of our interest, 
the ground state (with energy $E_0$) of the model \eqref{eq:H} is
in the sector $(S^z=0,q=0,P=+1,T=+1)$ when $L$ is even. 
In the TLL phase, the first excited state (with energy $E_{\rm S}$) 
is in the sector $(S^z=\pm 1,q=0, P=+1)$. 
In the N\'eel and dimer phases, 
pseudo ground states 
(with energies $E_{\rm N}$ and $E_{\rm D}$)
appear in the sectors $(S^z=0,q=\pi,P=-1,T=-1)$ and
$(S^z=0,q=\pi,P=+1,T=+1)$, respectively,
reflecting the two-fold ground-state degeneracy in the thermodynamic limit.
One can expect that the transitions between the three phases 
can be detected by observing crossing of these energy levels. 

\begin{figure}
\begin{center}
\includegraphics[width=0.5\textwidth]{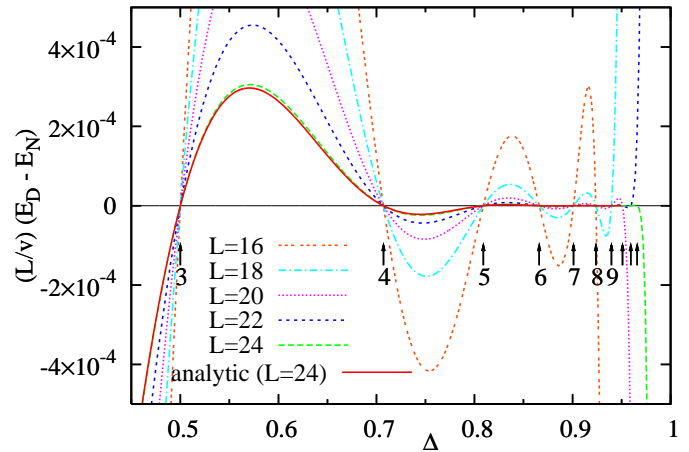}
\end{center}
\caption{
(color online) 
Energy difference $(L/v)(E_\mathrm{D}-E_\mathrm{N})$
in the unfrustrated case $J_2=0$. 
The arrows indicate the ``special'' points in Eq.~\eqref{eq:XXZ_special} 
with $n=3,4,\dots$, 
where $\lambda$ changes the sign. 
}
\label{fig:nldm_jt0}
\end{figure}

This expectation can be justified by
a detailed analysis of the finite-size spectra 
of the sine-Gordon theory.~\cite{Nomura95} 
See Appendix for details.
Here we briefly mention the basic idea.
The excitation energy $\Delta E_{\rm S}:=E_{\rm S}-E_0$
is associated with the operator 
$S_j^\pm \sim e^{\pm i\sqrt{\pi}\theta(x)}$. 
The excitation energies $\Delta E_{\rm N}:=E_{\rm N}-E_0$ and
$\Delta E_\mathrm{D}:=E_\mathrm{D}-E_0$
are associated with the order operators 
$(-1)^j S_j^z \sim \cos(\sqrt{4\pi}\phi)$ and 
$(-1)^j (S_j^+S_{j+1}^-+ \rm{h.c.}) \sim \sin(\sqrt{4\pi}\phi)$,
respectively. 
From the correspondence between the spectra and the operators, one obtains 
\begin{subequations}\label{eq:spectroscopy}
\begin{align}
&\Delta E_\mathrm{S} =
 \frac{2\pi v}{L} \left(\frac12-\frac{y(l)}4 \right),
 \label{eq:spectro1} \\
&\Delta E_\mathrm{N} =
 \frac{2\pi v}{L} \left(\frac12+\frac{y(l)}4-\frac{y_\phi(l)}2 \right),
 \label{eq:spectro2} \\
&\Delta E_\mathrm{D} =
 \frac{2\pi v}{L} \left(\frac12+\frac{y(l)}4+\frac{y_\phi(l)}2 \right),
 \label{eq:spectro3} 
\end{align}
\end{subequations}
where $l$ is related to the system size $L$ by $e^l=L/2\pi$. 
One can see that the crossing of $E_{\rm S}$ and $E_{\rm N}$ occurs 
when $y(l)=y_\phi(l)$, corresponding to the TL-N\'eel transition
in Fig.~\ref{fig:sineGordon}. 
Similarly, $E_{\rm S}= E_{\rm D}$ corresponds to the TL-dimer transition. 
Using the formula $L(E_{\rm D}-E_{\rm N})=2\pi v y_\phi(l)$, 
one can observe the running coupling constant, in particular, its sign change. 
In Fig.~\ref{fig:nldm_jt0}, $(L/v) (E_\mathrm{D}-E_\mathrm{N})$ is plotted 
for the $J_1$ chain ($J_2=0$). 
For a system of $L$ spins, we observe zeros of $y_\phi(l)$ 
at ``special'' points in Eq.~\eqref{eq:XXZ_special} up to $n=L/2$.
In the limit $L\to\infty$, the energy difference asymptotically
obeys\cite{Bortz09}
\begin{equation}
 \frac{L}{v} (E_\mathrm{D}-E_\mathrm{N}) =
 2\pi \lambda_0 \left( \frac{L}{2\pi} \right)^{2-2/\eta}. 
\end{equation}
This formula is also plotted for $L=24$ 
(red solid line in Fig.~\ref{fig:nldm_jt0}), 
which agrees well with the numerical result for $\Delta\lesssim 0.8$. 
For larger $\Delta$, better agreement is expected to be seen at larger $L$. 



\begin{figure}[t]
\begin{center}
 \includegraphics[width=0.5\textwidth]{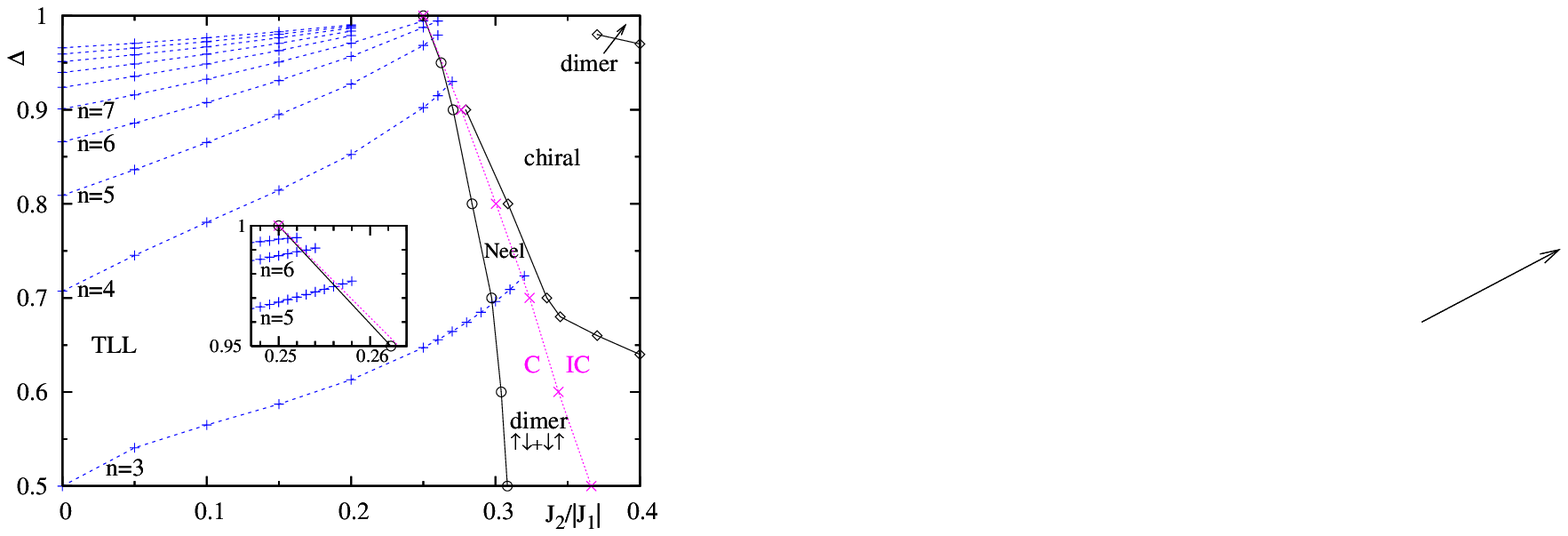}
\end{center}
\caption{
(color online)
Phase diagram of the model \eqref{eq:H}. 
The level spectroscopy has been performed for $L=24$. 
We have observed that finite-size effects are quite small
and can be neglected. 
The inset shows a zoom around the highly-degenerate point
$(J_2/|J_1|,\Delta)=(1/4,1)$.
}
\label{fig:phase}
\end{figure}

\section{Phase diagram}

The phase diagram is determined by the level spectroscopy and
the iTEBD analysis as presented in Fig.~\ref{fig:phase}. 
The blue ``+'' symbols show Gaussian fixed points determined
from the condition
$E_{\rm N} = E_{\rm D}$.   
The black open circles interpolated by lines show the transition
points from the TLL phase to the N\'eel
and dimer phases, which are
determined by $E_{\rm S}=E_{\rm N}$ and $E_{\rm S}=E_{\rm D}$, respectively.
This phase boundary between the TLL phase and the gapped ordered phases
is smoothly connected to the ferromagnetic transition
point\cite{Bursill95} $J_2/|J_1|=1/4$ in the isotropic case $\Delta=1$, 
at which the ground states are known to be highly degenerate.\cite{Hamada88} 
The Gaussian fixed lines with $\lambda=0$ start from the ``special'' points
given in Eq.~\eqref{eq:XXZ_special} at $J_2=0$, 
and are therefore labeled by $n=3,4,\dots$ for convenience.
Some of them continue even after the cosine potential becomes relevant
for $J_2/|J_1|>1/4$. 
From calculation of $L=24$ system, we have identified five lines
(corresponding to $n=3,\ldots,7$) extending into the parameter region
$J_2/|J_1|>1/4$ and $\Delta\lesssim 0.995$. 
Across these lines, there occur successive transitions
between the N\'eel and dimer phases 
as the anisotropy parameter $\Delta$ is changed toward unity. 
However, for $0.995\lesssim \Delta \le 1$,
Lanczos diagonalization does not converge well, 
due to highly degenerate nature around $(J_2/|J_1|,\Delta)=(1/4,1)$.
For this reason we have not been able to find the transition lines
beyond $n=7$.

Inside the N\'eel and dimer phases,
there is a Lifshitz line (pink ``$\times$'' symbols), 
where the short-range spin correlation in the $xy$ plane changes
its character from commensurate to incommensurate (indicated by C and IC in Fig.~\ref{fig:phase}). 
This line was determined by observing the peak position of the equal-time spin structure factor 
calculated by iTEBD.\cite{FSO} 
The level spectroscopy presented above is valid on the left side of this line.
By further increasing $J_2/|J_1|$, the system enters the vector chiral
ordered phase in which the vector product of neighboring spins, 
$\kappa_{j,j+1}^z:=\langle (\bm{S}_j\times\bm{S}_{j+1})^z\rangle$, 
is long-range ordered. 
The boundaries above and below the vector chiral phase (black diamond symbols) were determined with reasonable accuracy 
by observing the rapid increase of the order parameter $\kappa_{j,j+1}^z$ calculated by iTEBD.\cite{FSO} 
For $0.9\lesssim\Delta<1$ and $0.25<J_2/|J_1|\lesssim0.35$, 
we have not been able to locate the transition lines accurately 
because of a relatively poor convergence of iTEBD around the highly degenerate point $(J_2/|J_1|,\Delta)=(1/4,1)$. 
However, we expect that both the lines should continue to the point $(J_2/|J_1|,\Delta)=(1/4,1)$. 
The vector chiral phase is smoothly connected to that 
found in the frustrated ferromagnetic chain in a low magnetic field.\cite{Hikihara08,Sudan09,Heidrich09}  
Similar vector chiral phases have also been found previously 
in the antiferromagnetic $J_1$-$J_2$ model with easy-plane anisotropy\cite{Nersesyan98,Hikihara01}
or in a magnetic field.\cite{Kolezhuk05,McCulloch08,Okunishi08,Hikihara10}



The nature of the dimer phase at $\Delta\lesssim 0.7$ is easy
to understand,
as this phase extends to the XY point $\Delta=0$,
where the sign of $J_1$ can be reversed 
by performing the $\pi$ rotations around the $z$ axis of the spins
on every second sites.
From the fact that the ground state at the Majumdar-Ghosh point
$J_2=J_1/2>0$ and $\Delta=0$
(and more generally for all $\Delta>-1/2$)\cite{Majumdar69}
is given exactly by the product of singlet dimers 
$(|\!\ua\da\rangle-|\!\da\ua\rangle)/\sqrt{2}$,
one finds, through the above $\pi$-rotation transformation,
that the ground states at $J_2=-J_1/2>0$ and $\Delta=0$ 
are given by the dimer states 
whose dimer unit is now replaced by the triplet state 
$(|\!\ua\da\rangle+|\!\da\ua\rangle)/\sqrt{2}$.\cite{Chubukov91}
We expect that such a ``triplet'' dimer nature should survive away from the
XY point and define the order parameter of the dimer phase.
The nature of the N\'eel phase in $0.7\lesssim \Delta \lesssim 0.9$ 
is more elusive and will be discussed in detail in the next section.

\begin{figure}
\begin{center}
\includegraphics[width=0.5\textwidth]{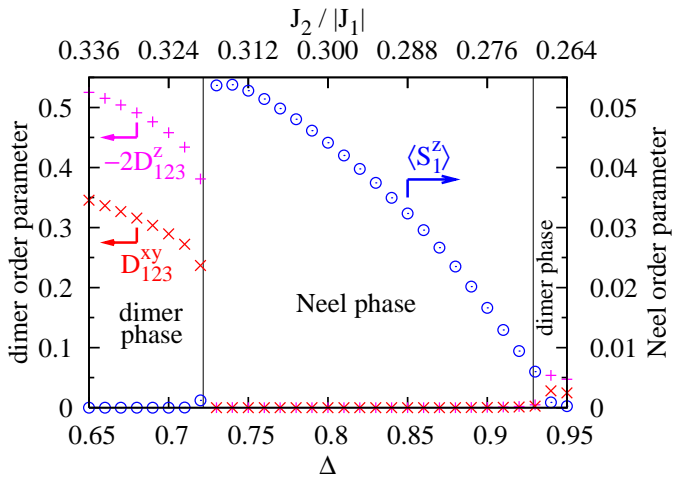}
\end{center}
\caption{
(color online) 
The dimer order parameters $D_{123}^{xy}$ and $-2D_{123}^z$ (left axis) 
and the local magnetization $\langle S_1^z \rangle$ (right axis),  
along the line connecting $(J_2/|J_1|,\Delta)=(0.336,0.65)$
and $(0.264,0.95)$, 
which approximately traces the Lifshitz line. 
The order parameters take on relatively large values along this line. 
The calculation was done by the iTEBD with the Schmidt rank $\chi=300$.
The vertical lines indicate the phase boundaries determined
by the level spectroscopy. 
}
\label{fig:order}
\end{figure}

So far our argument for the emergence of the N\'eel and dimer phases 
has been based on the effective SG theory. 
To confirm their existence in an independent and unbiased way, 
we have calculated the order parameters of these phases
using the iTEBD algorithm,\cite{Vidal07} 
which can address physical quantities in the thermodynamic limit directly 
through the use of the matrix product representation of (ground) states. 
When this algorithm is performed in an ordered phase, 
a variational state finally converges to a symmetry-broken state 
with a finite spontaneous order parameter 
(if it is allowed by the periodicity of the matrix product state). 
Hence, the N\'eel phase is detected by a local magnetization 
$\langle S_1^z \rangle$,  
and the dimer phase by dimer order parameters: 
\begin{eqnarray}
 D_{123}^{xy} &=& (S_1^x S_2^x + S_1^y S_2^y) - (S_2^x S_3^x + S_2^y S_3^y),\\
 D_{123}^z    &=& S_1^z S_2^z - S_2^z S_3^z.
\end{eqnarray}
These order parameters show alternating signs along the spin chain. 
We choose the site labellings in such a way that
$\langle S_1^z \rangle>0$ and $D_{123}^z<0$. 
We calculated these order parameters
for the parameter points $(J_2/|J_1|,\Delta)$ varying
approximately along the Lifshitz line; see Fig.~\ref{fig:order}.
In this figure we observe two transitions between the dimer and N\'eel phases, 
in agreement with the level spectroscopy analysis. 
In the dimer phases, $D_{123}^{xy}$ and $D_{123}^z$ show opposite signs 
in accord with the ``triplet'' nature of the dimers. 
[Notice that $D_{123}^{xy}=-2D_{123}^z=1/2$ in the exact ``triplet'' dimer
state at $(J_2/|J_1|,\Delta)=(1/2,0)$.] 

The sudden changes of the order parameters
in Fig.~\ref{fig:order} 
seem to indicate first-order nature of the transitions
between N\'eel and dimer phases for this parameter regime. 
On the contrary, a continuous transition of Gaussian type is expected
in the SG model (Fig.~\ref{fig:sineGordon}).
This discrepancy may be reconciled by considering the effect of
a higher-frequency cosine potential 
$\cos (8\sqrt{\pi}\phi)$,
which was ignored in the SG theory but is allowed by symmetry,
and can become relevant deep inside
the ordered phases.
With a negative coefficient,
the potential $\cos(8\sqrt{\pi}\phi)$
has four minima corresponding to the N\'eel and dimer orderings 
with doubly-degenerate ground states each. 
Different signs of $\lambda$ select different types of orderings, 
and thus the first-order transition at $\lambda=0$ separates the two phases. 


We finally note that, in the small region 
near the upper right corner of Fig.~\ref{fig:phase}, 
a weak dimerization with the {\it same} sign in $D_{123}^{xy}$ and $D_{123}^z$ was observed.\cite{FSO} 
In particular, $D_{123}^{xy}=2D_{123}^z$ in the isotropic case $\Delta=1$. 
This result indicates that this region is characterized as a dimer ordered phase 
having a distinct nature from the ``triplet'' dimer phase in $\Delta\lesssim 0.7$.

\section{Nature of the N\'eel phase}

\begin{figure}
\begin{center}
\includegraphics[width=0.5\textwidth]{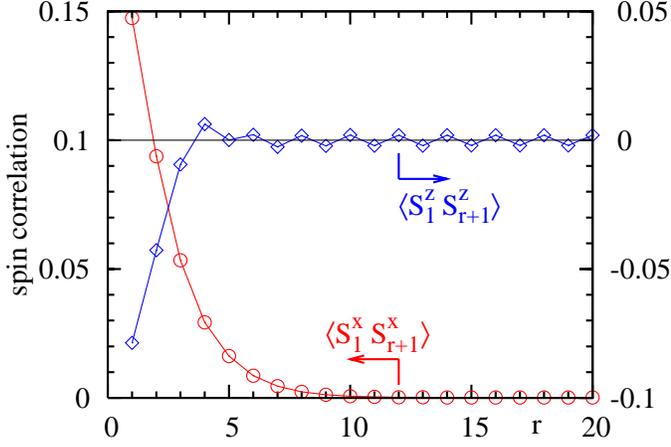}
\end{center}
\caption{
(color online)
Spin correlation functions 
$\langle S^x_1 S^x_{r+1} \rangle$ (left axis) and
$\langle S^z_1 S^z_{r+1} \rangle$ (right axis) 
for $J_2/|J_1|=0.31$ and $\Delta=0.74$, 
where the N\'eel order parameter is relatively large: 
$\langle S^z_1 \rangle = - \langle S^z_2 \rangle=0.04432$. 
}
\label{fig:corr}
\end{figure}

The appearance of the N\'eel phase is a natural consequence of
the SG theory \eqref{eq:sineGordon} with $\lambda>0$, 
but is quite counterintuitive in the presence of the easy-plane
nearest-neighbor ferromagnetic coupling. 
In Fig.~\ref{fig:order}, the N\'eel order parameter is one order of
magnitude smaller than the value $\langle S_1^z \rangle=1/2$ of
the pure N\'eel state $|\!\ua\da\ua\da\dots\rangle$, 
which indicates that the ground state is not well approximated
by such a simple product state. 
Here we analyze the nature of this phase in detail using iTEBD
with Schmidt rank $\chi=200$. 
We first analyze the spin correlations; see Fig.~\ref{fig:corr}.
We observe that 
the N\'eel character of the correlations appears only
at a relatively long-distance scale $r\gtrsim 6$. 
At a short-distance scale, $\langle S^x_1 S^x_{1+r} \rangle$ takes
larger amplitudes than $\langle S^z_1 S^z_{1+r} \rangle$,
reflecting the easy-plane anisotropy.
Furthermore, we note that $\langle S^z_1 S^z_3 \rangle$ takes
relatively large negative value, 
which is unusual in a N\'eel phase.

\begin{figure}
\begin{center}
\includegraphics[width=0.5\textwidth]{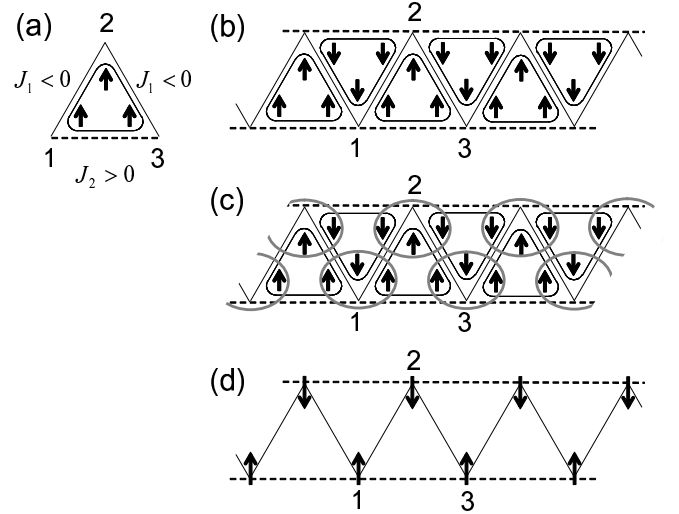}
\end{center}
\caption{
(a): Three-spin model. 
(b), (c), (d): Neel order from the trimer picture. 
The arrows show the $z$-component of the local magnetization.
}
\label{fig:NeelTrimer}
\end{figure}

\begin{figure}
\begin{center}
\includegraphics[width=0.5\textwidth]{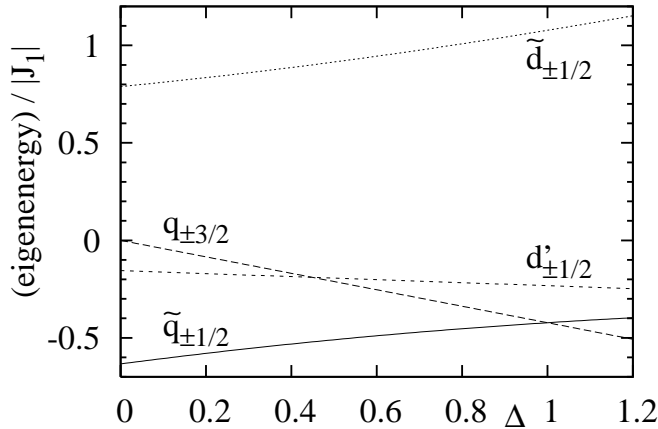}
\end{center}
\caption{
Spectrum of the three-spin model in Fig.~\ref{fig:NeelTrimer}(a) 
for $J_2/|J_1|=0.31$. 
The eigenstates are given in Eqs.~\eqref{eq:3spin_state}. 
}
\label{fig:spec_3spin}
\end{figure}

Such unusual correlations in a short-distance scale can be understood 
in terms of trimer degrees of freedom.
First, we consider a three-spin model shown in Fig.~\ref{fig:NeelTrimer}(a). 
In the isotropic case $\Delta=1$, 
the ground state is a quadruplet $\ket{q_\mu}$
(with $S^z_{\rm tot}=\mu=\pm\frac12,\pm\frac32$) when $J_2/|J_1|<1/2$. 
Because of the reflection symmetry around the site $2$, 
the excited states are classified into symmetric and antisymmetric doublets, 
$\{ \ket{d_\mu} \}$ and $\{ \ket{d_\mu^\prime} \}$ (with $\mu=\pm\frac12)$.
When the anisotropy $\Delta$ is introduced, 
$\ket{q_{\pm\frac12}}$ and $\ket{d_{\pm\frac12}}$ are mixed 
to form new eigenstates 
$\ket{\tilde{q}_{\pm\frac12}}$ and $\ket{\tilde{d}_{\pm\frac12}}$. 
The eigenstates are summarized as follows: 
\begin{subequations}\label{eq:3spin_state}
\begin{align}
 &\ket{q_{+\frac32}} =
 \ket{\!\ua\ua\ua}, \\
 &\ket{\tilde{q}_{+\frac12} (\gamma)} = \frac1{\sqrt{2+\gamma^2}} 
 (\ket{\!\ua\ua\da} + \ket{\!\da\ua\ua} + \gamma \ket{\!\ua\da\ua}), \\
 &\ket{\tilde{d}_{+\frac12} (\gamma)} = \frac1{\sqrt{4+2\gamma^2}} 
 (\gamma \ket{\!\ua\ua\da} + \gamma \ket{\!\da\ua\ua} -2 \ket{\!\ua\da\ua}),\\
 &\ket{d^\prime_{+\frac12}} = \frac1{\sqrt{2}}
 (\ket{\!\ua\ua\da} - \ket{\!\da\ua\ua}),  
\end{align}
\end{subequations}
with $\gamma=\frac12 (\sqrt{\tilde{\Delta}^2+8}-\tilde{\Delta})$ 
and $\tilde{\Delta}=\Delta+(J_2/|J_1|) (\Delta-1)$. 
Other eigenstates with $\mu=-1/2$ or $-3/2$ are obtained
by applying the spin reversal to the above. 
The eigenenergies are plotted in Fig.~\ref{fig:spec_3spin}. 
Under the easy-plane anisotropy $\Delta<1$, 
the ground states are given by $\ket{\tilde{q}_{\pm\frac12} (\gamma)}$ 
(with $\gamma>1$), 
in which $S_j^x$'s correlate ferromagnetically and
$S_j^z$'s antiferromagnetically
for any spin pair (if $\gamma<\sqrt2$). 
This property is in accord with the anomalous spin correlations 
in Fig.~\ref{fig:corr} for $r=1,2$. 
Note, however, that the local magnetization $\langle S_j^z \rangle$
is almost uniform 
in $\ket{\tilde{q}_{\pm\frac12} (\gamma)}$ as shown in Fig.~\ref{fig:NeelTrimer}(a), 
and we do not observe N\'eel ordering pattern at this level. 

Now we construct a N\'eel ordered state in a 1D chain as follows. 
(The procedure is analogous to the construction of the valence bond solid
ground state in the Affleck-Kennedy-Lieb-Tasaki model,\cite{Affleck87}
and more generally to the construction of a projected entangled-pair
state.\cite{Verstraete04})
The 1D model \eqref{eq:H} can be viewed as the model on the zigzag ladder, 
where $J_1$ and $J_2$ give the inter- and intra-chain couplings respectively.  
We place a trimer state $\ket{\tilde{q}_{+\frac12} (\gamma)}$ 
($\ket{\tilde{q}_{-\frac12} (\gamma)}$ resp.)\ on
every up (down resp.)\ triangle in the zigzag ladder,
as shown in Fig.~\ref{fig:NeelTrimer}(b). 
At this point, every site is shared by three neighboring trimers.
To define the state in the original Hilbert space, 
we bind the three spins at every site together to form a single spin-$1/2$, 
as indicated by an ellipse in Fig.~\ref{fig:NeelTrimer}(c). 
This is done by a projection operator 
$\ket{\!\ua} \bra{\psi_+} 
+ \ket{\!\da} \bra{\psi_-}$, 
where $\ket{\psi_\pm}$ are defined for the three spins in an ellipse. 
We require that (i) $\ket{\psi_\pm}$ have the reflection symmetry about
the central spin and that
(ii) they have sufficient overlap with
$\ket{\!\!\ua\da\ua}$ and $\ket{\!\!\da\ua\da}$, 
which are expected to have large weights in the state before projection
as can be seen in Fig.~\ref{fig:NeelTrimer}(c). 
The states satisfying (i) can be written in the same way as
in Eq.~(\ref{eq:3spin_state}a)-(\ref{eq:3spin_state}c). 
Considering also (ii), we set
$\ket{\psi_\pm}=\ket{\tilde{d}_{\pm\frac12}(\gamma^\prime)}$
with $\gamma^\prime$ appropriately tuned. 
The obtained state after the projection has a N\'eel ordering pattern
as shown in Fig.~\ref{fig:NeelTrimer}(d). 
We expect that this should give an approximate ansatz
(with two variational parameters, $\gamma$ and $\gamma^\prime$)
of the N\'eel ground state found in the present model. 



To support the trimer picture, 
we calculated the reduced density matrix $\rho_{123}$
for neighboring three spins $(1,2,3)$ 
from the ground state obtained by the iTEBD. 
It was done for the same parameter point as in Fig.~\ref{fig:corr}. 
Because of the reflection symmetry about the site $2$, 
$\rho_{123}$ is diagonalized in the form
\begin{equation}
\begin{split}
 \rho_{123} =
&\, \ket{       q _{+\frac32}          } a_+
  \bra{       q _{+\frac32}          }
 + \ket{\tilde{q}_{+\frac12}(\gamma_+)} b_+
   \bra{\tilde{q}_{+\frac12}(\gamma_+)} \\
&+\ket{\tilde{d}_{+\frac12}(\gamma_+)} c_+
  \bra{\tilde{d}_{+\frac12}(\gamma_+)}
 +\ket{       d _{+\frac12}^\prime   } e_+
  \bra{       d _{+\frac12}^\prime   } \\
&+ (\text{``}+\text{''} \to \text{``}-\text{''}), 
\end{split}
\end{equation}
where 
$\gamma_+=1.225$ and $\gamma_-=1.250$ are not far from 
$\gamma=1.122$ obtained in the three-spin problem. 
Large weights are found in 
$b_+=0.5803$, $b_-=0.2633$, $e_-=0.0999$ and $a_-=0.0472$, 
and the others are orders of magnitude smaller. 
We see that the trimer states $\ket{\tilde{q}_{\pm \frac12}(\gamma_\pm)}$ 
occupy about 84\% of the total weight.  

We also applied the systematic method for determining the order parameter 
proposed in Ref.~\onlinecite{FMO06}. 
In this method, we use the pair of symmetry-broken ground states
$|\Psi_{1,2}\rangle$ 
-- one is obtained directly by iTEBD and the other by applying
the spin reversal to it. 
For a subregion $A$ of the system, 
we construct the reduced density matrices $\rho_A^{1,2}$ from these states, 
and measure the mutual distance:
\begin{equation}
 {\rm diff}(\rho_A^1,\rho_A^2) = \sum_j |\lambda_j|,
\end{equation}
where $\lambda_j$'s are the eigenvalues of $\rho_A^1-\rho_A^2$. 
This measure takes a value from zero to two and 
shows to what extent the two states are distinguished in the concerned region. 
[See Refs.~\onlinecite{FMO06,Furukawa07} for details.] 
For one- and two-site regions, we obtain the same value 
${\rm diff}=4\langle S^z_1\rangle = 0.1773$. 
The three-site region $A=\{1,2,3\}$ gives a much larger value
${\rm diff}=0.9128$. 
These results indicate that, although the symmetry breaking can be detected
through a single-spin operator as expected for a N\'eel phase,
it is much more visible in terms of a three-spin operator. 
This also supports the trimer picture for the N\'eel state. 

Finally, we note that a N\'eel ordering 
in the presence of easy-plane anisotropy 
has been found in an antiferromagnetic model with {\em explicit} trimerization.\cite{Okamoto02} 
This example also suggests that the formation of trimers can change
the nature of the anisotropy of the system.

\section{Conclusions}

We have analyzed the phase diagram of the spin-$\frac12$ frustrated
ferromagnetic chain \eqref{eq:H}, 
starting from the bosonization picture in the unfrustrated case $J_2=0$. 
A crucial observation is that the effective sine-Gordon theory 
shows a cascade of sign changes in the coupling of the cosine potential
when changing the anisotropy $\Delta$. 
When the cosine potential is made relevant by
the competing $J_2$ exchange interaction,
it gives rise to alternate appearance of N\'eel and dimer orders. 
Although the N\'eel phase has a long-ranged staggered spin ordering pattern, 
we have argued that this ordering should rather be interpreted as
the ordering of the trimer degrees of freedom. 

The successive sign changes in the coupling of the cosine term
in the effective theory of the FM $J_1$ XXZ chain is explicitly seen
in the exact expression \eqref{eq:coupling_Lukyanov}
obtained in Refs.~\onlinecite{Lukyanov98,Lukyanov03}.
Our study has shown for the first time that
it has a dramatic physical consequence,
the alternation of the N\'eel and dimer ordered phases. 
It will be interesting to further examine the effect of the sign change
in the presence of other types of perturbations. 


\acknowledgments

We are grateful to Shigeki Onoda and Tsutomu Momoi for fruitful discussions. 
This work was supported in part by Grants-in-Aid for Scientific Research
(No.~20046016 and No.~21740295) from MEXT, Japan. 

\appendix*
\section{Finite-size spectra in the sine-Gordon theory} 

In this appendix, following Cardy\cite{Cardy86} and
Bortz {\it et al.},\cite{Bortz09} 
we present a simple derivation of the excitation
energies \eqref{eq:spectroscopy}  
based on the perturbation theory from the Gaussian model. 
This is complementary to the derivation by Nomura\cite{Nomura95}
using the correlation functions. 

In a finite-size system of length $L$, the effective Hamiltonian is given by 
\begin{align}
H = & H_{\rm Gauss} + H_{\cos} \notag \\
= &
\int_0^L dx
\left\{
  \frac{v}2
  \left[
   K \left(\frac{d\theta}{dx}\right)^2
   + \frac1{K} \left(\frac{d\phi}{dx}\right)^2
  \right]
 \right.\notag \\
& \left.\qquad\qquad
  - \frac{v\lambda}{2\pi} \cos (\sqrt{16\pi}\phi) \right\},
\end{align}
where the lattice spacing is set to unity. 

Let us start from the Gaussian part $H_{\rm Gauss}$. 
It is useful to rescale the fields as $\Phi\equiv \phi/\sqrt{K}$
and $\Theta\equiv \sqrt{K}\theta$ 
to simplify the Hamiltonian: 
\begin{equation}
 H_{\rm Gauss} = 
 \int_0^L dx~
  \frac{v}2 \left[
    \left(\frac{d\Theta}{dx}\right)^2 + \left(\frac{d\Phi}{dx}\right)^2
  \right].
\end{equation}
Assuming the periodic boundary condition,
we perform the mode expansions of the bosonic fields:
\begin{align}
\label{eq:expand_Phi}
&\Phi(x) = \Phi_0 + \widetilde{Q}\frac{x}L 
+ \sum_{n\ne 0} \frac{e^{-|n|/2\Lambda}}{\sqrt{4\pi|n|}}
 \left(e^{ik_n x}a_n + e^{-ik_n x}a_n^\dagger \right),\\
\label{eq:expand_Theta}
&\Theta(x) = \Theta_0 + Q \frac{x}L \nonumber \\
&\hspace{1cm}
 + \sum_{n\ne 0} \frac{e^{-|n|/2\Lambda}}{\sqrt{4\pi|n|}}
  {\rm sign}(n) \left(e^{ik_n x}a_n + e^{-ik_n x}a_n^\dagger \right),
\end{align}
with $k_n=2\pi n/L$. 
Here, $a_n$ with $n>0$ ($n<0$) represents a right (left) moving mode. 
We have introduced an ultraviolet cutoff $\Lambda$, 
which will be determined later. 
The operators in the expansions obey the commutation relations
\begin{equation}
\begin{split}
&[\Phi_0,\Theta_0]=-i,
 \quad
[\widetilde{Q},\Theta_0]=i,
 \quad
[Q,\Phi_0]=i,\\
&[a_n,a_m^\dagger]=\delta_{nm}. 
\end{split}
\end{equation}
Using the mode expansions, the Hamiltonian is diagonalized as 
\begin{equation}
 H_{\rm Gauss} = \frac{2\pi v}L \left[ 
   \frac1{4\pi} ( \widetilde{Q}^2 + Q^2 ) 
 + \sum_{n\ne 0} |n| a_n^{\dagger} a_n
 \right], 
\end{equation}
where $\Lambda$ is taken to infinity. 
The ground state $\ket{0}$ is defined by 
$a_n\ket{0} = \widetilde{Q}\ket{0} = Q\ket{0}=0$ 
and its energy is set to zero. 
The compactification radii $R$ and $\tilde{R}$ for $\Phi$ and $\Theta$
are given by
\begin{equation}
 2K = \frac{1}{2\pi R^2} = 2\pi \tilde{R}^2.
\end{equation} 
Then $\widetilde{Q}$ and $Q$ are quantized as
\begin{equation}
 \widetilde{Q} = S^z / \tilde{R}, \quad
 Q = m / R, \quad
 \text{with}~S^z,m \in \mathbb{Z}. 
\end{equation}
Here, the integer  $S^z$ coincides with the total magnetization. 
The entire spectrum is then given by
\begin{equation}
\begin{split}
 &\Delta E_0 (S^z,m,\{m_n\})\\
 &= \frac{2\pi v}L \left[ 
   \frac1{4\pi} \left( \frac{(S^z)^2}{\tilde{R}^2} + \frac{m^2}{R^2} \right) 
 + \sum_{n\ne 0}|n| m_n
 \right]
\end{split}
\end{equation}
with $m_n\ge 0$ integer,   
and the corresponding eigenstates are
\begin{equation}
\begin{split}
 \ket{ S^z,m,\{m_n\}}
 = e^{i S^z \Theta_0/\tilde{R} + i m \Phi_0 / R} 
   \prod_{n\ne 0} (a_n^{\dagger})^{m_n} \ket{0}.
\end{split}
\end{equation}
The excitation $\Delta E_{\rm S}$ is associated with
$e^{\pm i \Theta_0/\tilde{R}}$ 
and is thus given by 
\begin{equation}\label{eq:DE_S}
\Delta E_{\rm S} = \Delta E(\pm 1,0,0)
 = \frac{2\pi v}L \frac1{4\pi \tilde{R}^2}
 = \frac{2\pi v}L \frac1{2+y} ,
\end{equation}
with $y\equiv 4K-2$. 
This coincides with the expression \eqref{eq:spectro1} when $|y| \ll 1$. 
On the other hand, $\Delta E_{\rm N/D}$ are associated with 
$\cos (\Phi_0/R)$ and $\sin (\Phi_0/R)$ respectively. 
In the Gaussian model, these energies are degenerate and are given by 
\begin{eqnarray}
 \Delta E_{\rm N/D} =
 \Delta E(0,\pm 1,0)= \frac{2\pi v}L \frac1{4\pi R^2}= \frac{2\pi v}L K.
 \quad
\end{eqnarray} 

So far, we have ignored the cutoff $\Lambda$. 
It is determined so that the two-point correlation function of vertex
operators satisfies 
the simple normalization condition 
\begin{equation}\label{eq:norm_Phi}
\langle0|e^{i\mu\Phi(x)}e^{-i\mu\Phi(x^\prime)}|0\rangle
=\frac{1}{|x-x^\prime|^{\mu^2/2\pi}}
\end{equation}
for $|x-x^\prime|\ll L$. 
Using the expansion \eqref{eq:expand_Phi}, 
the l.h.s.\ of Eq.~\eqref{eq:norm_Phi} is evaluated as 
\begin{align}
& \exp \left[ -\mu^2 \bra{0} \Phi(x)\Phi(x^\prime) \ket{0} \right] \notag \\
& = \exp\!\left[-\mu^2\sum_{n\ne0}\frac{e^{-|n|/\Lambda}}{4\pi|n|}
  \left(1-e^{2\pi i|n|(x-x^\prime)/L}\right)\right] \notag\\
& = \left|\frac{1-e^{-\frac{1}{\Lambda}+i\frac{2\pi (x-x^\prime)}{L}}}
              {1-e^{-\frac{1}{\Lambda}}}\right|^{-\mu^2/2\pi} \notag\\
& \approx \left(2\Lambda\sin\frac{\pi|x-x^\prime|}{L}\right)^{-\mu^2/2\pi}.
\end{align}
Hence, we set $\Lambda=L/2\pi$ to satisfy Eq.~\eqref{eq:norm_Phi}.

Now we introduce $H_{\cos}$ as a perturbation. 
It splits the degeneracy between $\Delta E(0,\pm 1,0)$. 
We calculate the matrix element:
\begin{equation}
\begin{split}
 C &:= \bra{0,-1,0} H_{\cos} \ket{0,1,0}\\
  &= -\frac{v\lambda}{2\pi} \int_0^L dx\,
    \bra{0} e^{i\Phi_0/R} \cos\left( \frac{2\Phi(x)}{R} \right)
            e^{i\Phi_0/R} \ket{0}.
\end{split}
\end{equation}
Using the expansion \eqref{eq:expand_Phi}, the integrand is evaluated as 
\begin{equation}
\begin{split}
&
\langle0|
 e^{i\Phi_0/R} \cos\left(\frac{2\Phi(x)}{R}\right) e^{i\Phi_0/R}
|0\rangle \\
&= \frac{1}{2}\langle0|e^{i\Phi_0/R}e^{-2i\Phi(x)/R}e^{i\Phi_0/R}|0\rangle
\\
&=
\frac{1}{2}\langle0|
\exp\!\left[-\frac{2i}{R}\sum_{n\ne0}
\frac{e^{-|n|/2\Lambda}}{\sqrt{4\pi|n|}}
\left(e^{2\pi inx/L}a^{}_n
  + \mathrm{H.c.}
\right)
\right]\!
|0\rangle
\\
&=
\frac{1}{2}\exp\!\left(-\frac{1}{\pi R^2}\sum^\infty_{n=1}
\frac{e^{-n/\Lambda}}{n}\right)
= \frac12 \Lambda^{-4K}. 
\end{split}
\end{equation}
In first-order perturbation theory
the new eigenstates are given by the linear combinations 
\begin{equation}
 \frac1{\sqrt{2}} ( \ket{0,1,0} \pm \ket{0,-1,0} ),
\end{equation}
and the corresponding eigenenergies are 
\begin{equation} \label{eq:DE_ND}
\begin{split} 
 \Delta E_{\rm N/D} 
 = \frac{2\pi v}L K \pm C
 = \frac{2\pi v}L
 \left[ \frac12 + \frac{y}4 \mp \lambda \left(\frac{2\pi}{L}\right)^y \right]. 
\end{split}
\end{equation} 
Equations \eqref{eq:DE_S} and \eqref{eq:DE_ND} are non-perturbative
in $y$ and perturbative in $\lambda$. 

So far, we have used the bare coupling constants,
$y$ and $\lambda$. 
Now we relate the above results to the running coupling constants,
$y(l)$ and $y_\phi(l)$, 
focusing on the vicinity of the multicritical point $(y,y_\phi)=(0,0)$. 
Both the cosine potential and the change in the Luttinger parameter $y$
are marginal perturbations at this point. 
In general, the excitation energy $\Delta E_n$ associated
with an operator ${\cal O}_n$ 
is related to the inverse of the correlation length $\xi_n$ of the operator. 
Under a global scale transformation by a factor $e^l$, it scales as
\begin{equation}
 \Delta E_n \propto \xi_n^{-1} (y,y_\phi,L^{-1})
 = e^{-l} \xi_n^{-1} (y(l),y_\phi(l),L^{-1}e^l). 
\end{equation}
Setting $e^l=L/2\pi$, this is given by a universal function 
in terms of $y(l)$ and $y_\phi(l)$: 
\begin{equation}\label{eq:DEn_uni}
 \Delta E_n = \frac{2\pi}L \Phi_n (y(l),y_\phi(l)). 
\end{equation}
Namely, $L \Delta E_n$ depends on $L$ only through $y(l)$ and $y_\phi(l)$. 

We compare this with Eqs.~\eqref{eq:DE_S} and \eqref{eq:DE_ND}. 
As shown in Fig.~\ref{fig:sineGordon}, for $y>|y_\phi|$,
$y(l)$ finally converges to a constant. 
If the bare value of $y$ is sufficiently close to the final constant, 
the RG equations are solved as 
\begin{equation}
 y(l)=y, ~~ y_\phi(l)=y_\phi e^{-yl}. 
\end{equation}
Thus, under the correspondence $e^l=L/2\pi$, 
Eqs.~\eqref{eq:DE_S} and \eqref{eq:DE_ND} are expressed
by Eqs.~\eqref{eq:spectroscopy}, 
which show a linear dependence on $y(l)$ and $y_\phi(l)$. 
Equation \eqref{eq:DEn_uni} implies that
we can extend the range of validity of Eqs.~\eqref{eq:spectroscopy} 
to all $|y(l)|,|y_\phi(l)| \ll 1$.



\newcommand{\etal}{{\it et al.}}
\newcommand{\PRL}[3]{Phys.\ Rev.\ Lett.\ {\bf #1}, \href{http://link.aps.org/abstract/PRL/v#1/e#2}{#2} (#3)}
\newcommand{\PRLp}[3]{Phys.\ Rev.\ Lett.\ {\bf #1}, \href{http://link.aps.org/abstract/PRL/v#1/p#2}{#2} (#3)}
\newcommand{\PRA}[3]{Phys.\ Rev.\ A {\bf #1}, \href{http://link.aps.org/abstract/PRA/v#1/e#2}{#2} (#3)}
\newcommand{\PRAp}[3]{Phys.\ Rev.\ A {\bf #1}, \href{http://link.aps.org/abstract/PRA/v#1/p#2}{#2} (#3)}
\newcommand{\PRB}[3]{Phys.\ Rev.\ B {\bf #1}, \href{http://link.aps.org/abstract/PRB/v#1/e#2}{#2} (#3)}
\newcommand{\PRBp}[3]{Phys.\ Rev.\ B {\bf #1}, \href{http://link.aps.org/abstract/PRB/v#1/p#2}{#2} (#3)}
\newcommand{\PRBR}[3]{Phys.\ Rev.\ B {\bf #1}, \href{http://link.aps.org/abstract/PRB/v#1/e#2}{#2} (R) (#3)}
\newcommand{\PRBRp}[3]{Phys.\ Rev.\ B {\bf #1}, \href{http://link.aps.org/abstract/PRB/v#1/e#2}{R#2} (#3)}
\newcommand{\arXiv}[1]{arXiv:\href{http://arxiv.org/abs/#1}{#1}}
\newcommand{\condmat}[1]{cond-mat/\href{http://arxiv.org/abs/cond-mat/#1}{#1}}
\newcommand{\JPSJ}[3]{J. Phys.\ Soc.\ Jpn.\ {\bf #1}, \href{http://jpsj.ipap.jp/link?JPSJ/#1/#2/}{#2} (#3)}
\newcommand{\PTPS}[3]{Prog.\ Theor.\ Phys.\ Suppl.\ {\bf #1}, \href{http://ptp.ipap.jp/link?PTPS/#1/#2/}{#2} (#3)}
\newcommand{\hreflink}[1]{\href{#1}{#1}}







\begin{thebibliography}{99}
\bibitem{note_Sp}
Note that $J_1$ is negative in the present paper 
while it is positive in Refs.~\onlinecite{Giamarchi04,Gogolin98} and
other standard references. 
The sign of $J_1$ can be changed under the $\pi$-rotations of spins
around the $z$ axis on every second sites. 
The uniform and staggered parts in Eq.~\eqref{eq:Sp_boson} are
interchanged under this transformation.
\end{thebibliography}
\end{document}